\documentclass[aps,prb,preprint,superscriptaddress]{revtex4-2}

\usepackage{graphicx}        

\begin{document}

\title{In-situ observation of the formation of laser-induced periodic surface structures with extreme spatial and temporal resolution}
\author{K. Sokolowski-Tinten}
\affiliation{Faculty of Physics and Centre for Nanointegration Duisburg-Essen, University of Duisburg-Essen, Lotharstrasse 1, 47048 Duisburg, Germany}
\email{Klaus.Sokolowski-Tinten@uni-due.de}
\author{J. Bonse}
\affiliation{Bundesanstalt für Materialforschung und –prüfung (BAM), Unter den Eichen 87, 12205 Berlin, Germany}
\author{A. Barty}
\affiliation{Deutsches Elektronen-Synchrotron DESY, Notkestrasse 85, 22607 Hamburg, Germany}
\author{H. N. Chapman}
\affiliation{Center for Free-Electron Laser Science, Deutsches Elektronen-Synchrotron DESY, Notkestrasse 85, 22607 Hamburg, Germany}
\affiliation{Department of Physics and The Hamburg Centre for Ultrafast Imaging, Universit\"at Hamburg, Luruper Chaussee 149, 22761 Hamburg, Germany}
\author{S. Bajt}
\affiliation{Deutsches Elektronen-Synchrotron DESY, Notkestrasse 85, 22607 Hamburg, Germany}
\affiliation{The Hamburg Centre for Ultrafast Imaging, Universit\"at Hamburg, Luruper Chaussee 149, 22761 Hamburg, Germany}
\author{M. J. Bogan}
\affiliation{Lawrence Livermore National Laboratory, 7000 East Avenue, Livermore, CA 94551, USA}
\author{S. Boutet}
\affiliation{SLAC National Accelerator Laboratory, 2575 Sand Hill Road, Menlo Park, CA 94025, USA}
\author{A. Cavalleri}
\affiliation{Max Planck Institute for the Structure and Dynamics of Matter, Luruper Chaussee 149, 22761 Hamburg, Germany}
\author{S. D\"usterer}
\affiliation{Deutsches Elektronen-Synchrotron DESY, Notkestraße 85, 22607 Hamburg, Germany}
\author{M. Frank}
\affiliation{Lawrence Livermore National Laboratory, 7000 East Avenue, Livermore, CA 94551, USA}
\author{J. Hajdu}
\affiliation{Department of Cell and Molecular Biology, Uppsala Universitet, Box 596, 751 24 Uppsala, Sweden}
\affiliation{The European Extreme Light Infrastructure, Institute of Physics, Academy of Sciences of the Czech Republic, Za Radnici 835, 25241 Dolní Břežany, Czech Republic}
\author{S. Hau-Riege}
\affiliation{Lawrence Livermore National Laboratory, 7000 East Avenue, Livermore, CA 94551, USA}
\author{S. Marchesini}
\affiliation{SLAC National Accelerator Laboratory, 2575 Sand Hill Road, Menlo Park, CA 94025, USA}
\author{N. Stojanovic}
\affiliation{Deutsches Zentrum für Luft- und Raumfahrt (DLR) Berlin-Adlersdorf, Rutherfordstrasse 2, 12489 Berlin, Germany}
\author{R. Treusch}
\affiliation{Deutsches Elektronen-Synchrotron DESY, Notkestraße 85, 22607 Hamburg, Germany}

\begin{abstract}
Irradiation of solid surfaces with intense ultrashort laser pulses represents a unique way of depositing energy into materials. It allows to realize states of extreme electronic excitation and/or very high temperature and pressure, and to drive materials close to and beyond fundamental stability limits. As a consequence, structural changes and phase transitions often occur along unusual pathways and under strongly non-equilibrium conditions. Due to the inherent multiscale nature – both temporally and spatially – of these {\it irreversible} processes their direct experimental observation requires techniques that combine high temporal resolution with the appropriate spatial resolution and the capability to obtain good quality data on a single pulse/event basis. In this respect fourth generation light sources, namely short wavelength, short pulse free electron lasers (FELs) are offering new and fascinating possibilities. As an example, this chapter will discuss the results of scattering experiments carried at the FLASH free electron laser at DESY (Hamburg, Germany), which allowed us to resolve laser-induced structure formation at surfaces on the nanometer to sub-micron length scale and in temporal regimes ranging from picoseconds to several nanoseconds with sub-picosecond resolution.
\end{abstract}

\date{\today}
\maketitle

\section{Introduction}
\label{intro}
Irradiation of solid surfaces with intense ultrashort laser pulses represents a unique way of depositing energy into materials. It allows to realize states of extreme electronic excitation and/or very high temperature and pressure, and to drive materials close to and beyond fundamental stability limits. As a consequence, structural changes and phase transitions often occur along unusual pathways and under strongly non-equilibrium conditions \cite{rethfeld04}.

Moreover, such laser-solid interactions at high intensities and fluences are not only interesting from a fundamental physics point of you, but represent also the basis for numerous applications of lasers for materials processing and -synthesis (e.g.\ see Part 3 of this book and references therein). In this general context the formation of so-called {\bf L}aser-{\bf I}nduced {\bf P}eriodic {\bf S}urface {\bf S}tructures (LIPSS) or ripples represent an interesting and very common case under typical processing conditions, namely irradiation of {\it real} (i.e.\ rough) surfaces with multiple/many pulses.

LIPSS have been first reported in 1965 by Birnbaum \cite{birnbaum65} on semiconductors irradiated with pulses from a free-running ruby-laser ($\approx$ 200\,$\mu$s duration). Since then the topic has attracted continuous interest, not only to understand the basic mechanisms that lead to these sub-micrometer to nanoscale surface modifications, but also because LIPSS are considered as a versatile tool to achieve nanoscale structuring/patterning and functionalization of surfaces for a variety of applications (see \cite{florian20} and references therein, as well as Chapter 23 of this book by Mezera et al.).

LIPSS have been observed on almost all kinds of materials and with any kind of laser. Depending on the LIPSS periodicity different mechanisms are discussed in the very extensive literature. Meanwhile accepted is the view that LIPSS with a periodicity of the order of the laser wavelength (so-called low spatial frequency LIPSS: LSFL) can be attributed to the interference between the incident laser wave and waves scattered/excited at the surface leading to an intensity modulation and thus to a (periodically) modulated energy deposition \cite{emmony73}. In 1983 Sipe et al.\ \cite{sipe83} have developed this picture into a rigorous electromagnetic model which is frequently used and has been further developed since then\cite{dufft09,skolski10,hoehm12}.

More recently and so far only after irradiation with multiple ultrashort laser pulses LIPSS with a periodicity much smaller than the laser wavelength have been observed. The mechanisms which lead to these high spatial frequency LIPSS (HSFL) are not fully clear yet. HSFL have been for example attributed to near-field scattering and interference effects \cite{rudenko17}, in conjunction with surface second harmonic generation (e.g.\ \cite{jia05, guo08, dufft09}), but also self-organization processes have been proposed \cite{costache04,reif08}. Very recently, a coupled electromagnetic-hydrodynamic mechanism was proposed, where thermocapillary melt flows (Maragoni convection instability) that are initiated by spatial gradients caused by polarization-dependent absorption of random nanobumps and nanoholes cause the formation of HSFL \cite{rudenko20}.

However, it is not the purpose of this contribution to discuss in detail all models and controversies about LIPSS (for reviews of the current understanding the reader is referred to \cite{bonse17, bonse20, bonse20b} and references therein), but to reveal and emphasize the {\it dynamical} aspects of LIPSS-formation. In the vast majority of published work LIPSS have been studied {\it post mortem} through an analysis of the permanent modifications of the irradiated surface. A comparatively small number of time-resolved experiments addressing both, the nanosecond \cite{ehrlich82,keilmann83,young83a,young84,lee92} as well as the femtosecond and picosecond time range \cite{hoehm13, murphy13, jia14, kafka15, garcia16, zhou17, liu18}, have been performed. Using optical probing direct imaging \cite{murphy13, jia14, kafka15, garcia16, zhou17, liu18} as well as diffraction/scattering techniques \cite{ehrlich82,keilmann83,young83a,young84,lee92, hoehm13} have been applied. However, the spatial resolution was limited by the optical probe wavelength which made it difficult/impossible to follow the transient structural evolution of the laser-irradiated surface on sub-micrometer to nanometer length scales. For a more general discussion on time-resolved optical probing the reader is referred to Chapter 7 (M. Lechuga et al.) of this book.

Here we present results of time-resolved scattering experiments performed at the extreme ultraviolet (XUV) free electron laser (FEL) FLASH at DESY in Hamburg (Germany) \cite{ayvazyan06}. The unique combination of short wavelength, ultrashort pulse duration, and high photon flux has opened up new and exciting possibilities since it allows the investigation of {\it irreversible} processes with high temporal resolution and the appropriate nanoscale spatial resolution. We make use of these possibilities to study in the time domain the formation of LIPSS at the surface of laser-excited thin Silicon films \cite{note1}.

\section{Time-resolved XUV-scattering at FLASH}
\label{flash}
 The FLASH free electron laser at DESY (Hamburg/Germany) was the first FEL operating at wavelengths below 50\,nm \cite{ayvazyan06}. It has been available to users since 2005 serving a broad scientific community interested in atomic, molecular, and cluster physics, high energy density research, dynamics at surfaces surface, and diffraction imaging with high spatial resolution (see \cite{bostedt09, rossbach19} and references therein).

\begin{figure}[htb]
\centering
\includegraphics[width=15cm]{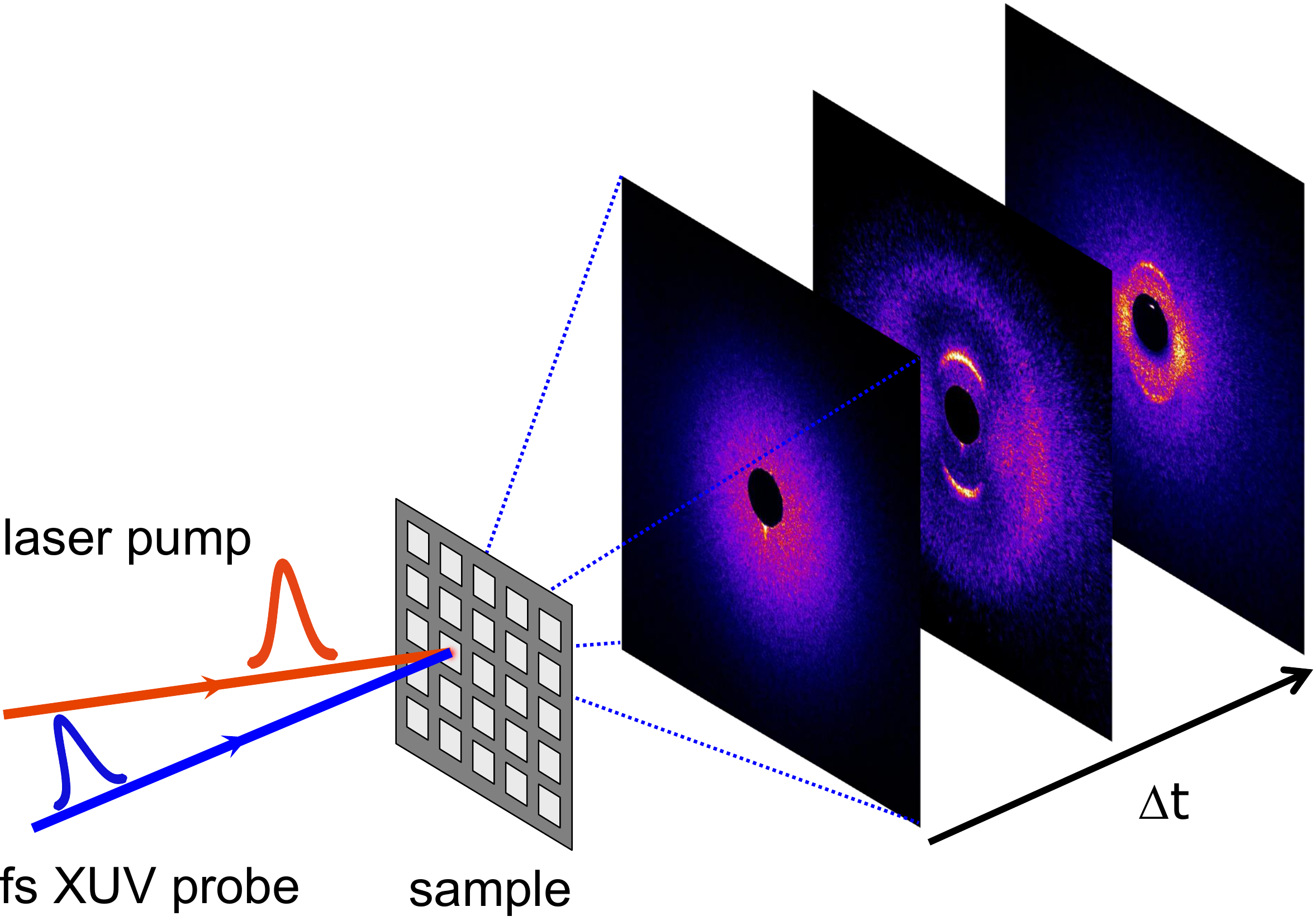}
\caption{\label{scheme1} Schematic of the time-resolved single-pulse scattering experiment performed at the XUV free electron laser FLASH.}
\end{figure}

 To address the particular topic of this contribution, namely the dynamics of nanoscale structure formation at laser-irradiated surfaces, we carried out a series of time-resolved laser-pump -- XUV scattering probe experiments to follow the laser-induced changes with femtosecond temporal and nanometer spatial resolution. A schematic of the experimental setup is shown in Fig.\ \ref{scheme1}.

 Experiments were performed at beamline BL2 \cite{tiedtke09} and FLASH was operated in the single-bunch ultrashort-pulse mode, providing pulses at a wavelength of 13.5\,nm, a pulse duration of 10--20\,fs and a mean pulse energy of about 20\,$\mu$J (corresponding to $>$10$^{12}$ XUV photons per pulse). These pulses were focused onto the sample under study (see below) to a spot size of approximately 20\,$\mu$m full width at half maximum (FWHM).

 Scattering experiments were carried out in a normal-incidence transmission geometry. A 45$\rm ^o$ graded multilayer mirror with a central hole was used to selectively collect the radiation scattered by the sample and direct it onto a XUV-sensitive CCD-detector while blocking at the same time any background scattering, plasma emission from the target etc., as well as the direct beam (which passed through the central hole). At $\rm \lambda _{XUV}$ = 13.5\,nm this diffraction setup allowed to measure scattering patterns over a spatial frequency range (spatial frequency $q = \left(2\pi /\lambda _{XUV}\right)\cdot sin(\theta)$; $\theta$ diffraction angle) from $\pm$10\,\,$\mu$m$^{-1}$ (determined by the size of the central hole) to $\pm$110\,$\mu$m$^{-1}$ (detector size), providing an effective spatial resolution of better than 50\,nm \cite{barty08}. For further details the reader is referred to \cite{bajt08, chapman06}.

 Samples comprised polycrystalline Silicon films with 100\,nm thickness deposited onto arrays of free-standing, 20\,nm thick Si$_3$N$_4$ membranes supported by a Silicon wafer frame. Such target arrays allowed replacement of the sample between consecutive exposures, because single-pulse irradiation with both, the optical laser as well as the FEL, lead to irreversible sample modifications (e.g.\ its destruction).

 These samples have been irradiated with 12\,ps, 523\,nm laser pulses \cite{redlin11} at an angle of incidence of 47$\rm ^o$ (normally p-polarized) and an incident fluence of about 1.7\,J/cm$^2$ (pump beamsize on the sample $30 \times 40\,\mu{\rm m^2}$ FWHM), well above the ablation threshold of $\lesssim\,{\rm 0.5\,J/cm}^2$. The delay time $\Delta t$ between the optical pump and the XUV probe was varied over a range from -50\,ps to 4.5\,ns with a help of a mechanical delay line.

 For each delay time 3 - 5 diffraction images have been recorded (each on a fresh sample window) and the data were averaged to increase the signal-to-noise ratio. Frequently detector images without pump and probe, pump only, and probe only have been recorded to allow for background correction as well as to characterize the scattering of the non-excited samples.

\section{Experimental Results}
\label{exps}
\begin{figure}[htb]
\centering
\includegraphics[width=12cm]{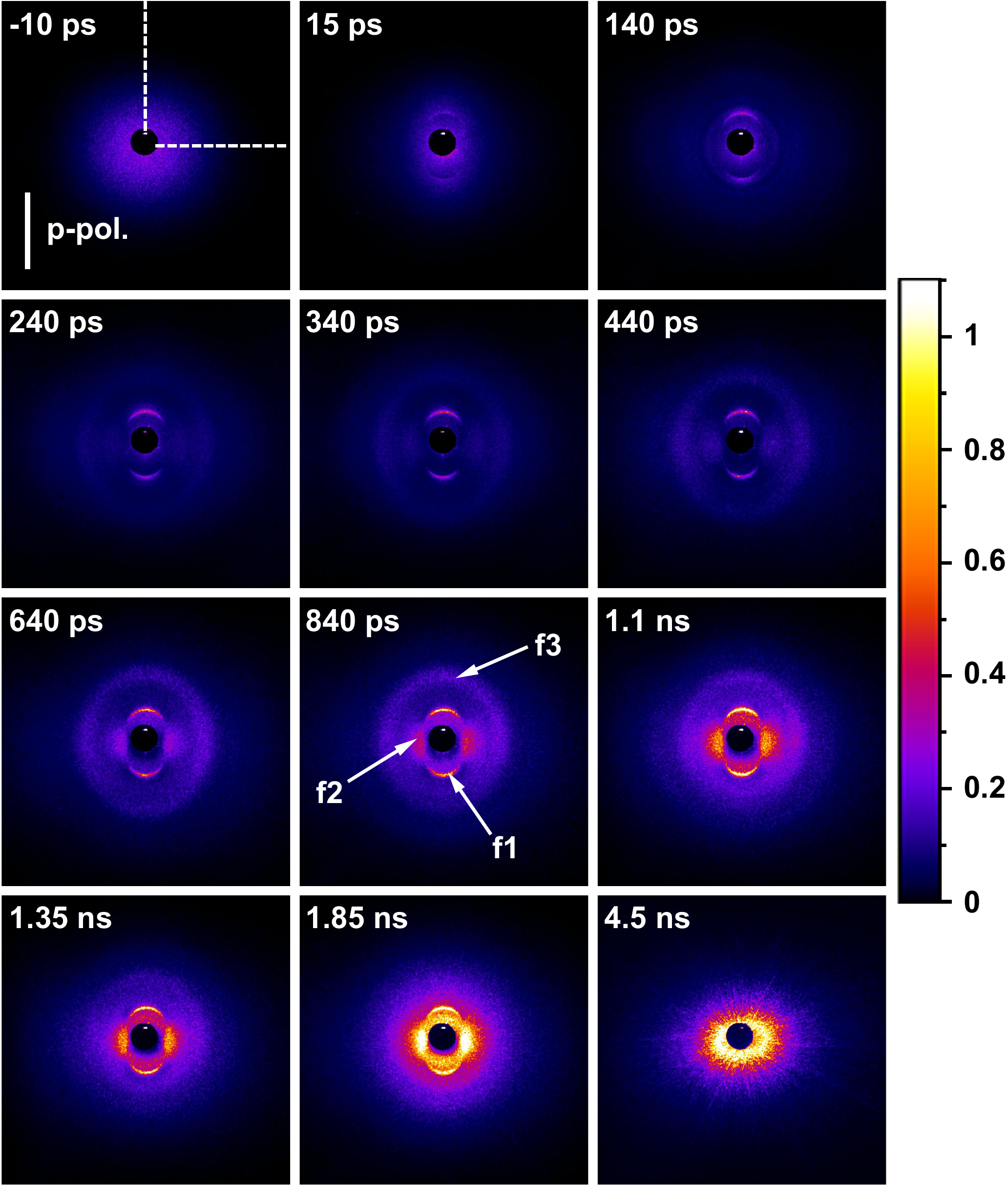}
\caption{\label{lipps_series} Transient scattering patterns (in false color representation) of a 100\,nm polycrystalline Si-film on a 20\,nm Si$_3$N$_4$ membrane (size: $100\,\times\,100\,\mu$m$^2$) as a function of delay time $\Delta t$ between the 12\,ps, 523\,nm optical pump pulse (fluence $\approx$ 1.7\,J/cm$^2$) and the 10-20\,fs, 13.5\,nm XUV probe pulse. Displayed spatial frequency range (horizontal and vertical): -92\,$\mu$m$^{-1}$ to +92\,$\mu$m$^{-1}$; intensity scale in arbitrary units, but the same for all frames; solid white bar in the first frame: direction of laser polarization (projection onto the surface).}
\end{figure}

As first results we present here data obtained on a 100\,nm thick polycrystalline Silicon film, deposited on $100\,\times\,100\,\mu$m$^2$ sized Si$_3$N$_4$ membrane windows. Fig.\ \ref{lipps_series} shows a sequence of transient scattering patterns for different pump-probe delay times. The intensity scale of the false-color representation is in arbitrary units, but the same for all frames/delays. The black disk in the center of each image represents the "beam stop" realized by the hole in the XUV multilayer mirror. The projection of the pump laser beam polarization (p-pol.) onto the sample surface is indicated by the vertical solid white bar in the first frame of Fig.\ \ref{lipps_series}.

Before discussing the temporal evolution of these scattering patterns we want to emphasize their "grainy" structure, which represent speckles due to the spatially coherent XUV illumination. From the average speckle size $\delta q_S \approx 0.4\,\mu {\rm m}^{-1}$ the diameter $d$ of the coherently illuminated sample area  can be estimated as $d = 2\pi/\delta q_S \approx 16\,\mu {\rm m}$, which agrees well with the nominal XUV spot size of $20\,\mu {\rm m}$ on the sample, highlighting the high spatial coherence of the FEL-radiation \cite{singer08}.

The diffraction pattern prior to excitation (-10\,ps delay) does reveal that the polycrystalline Silicon film has a transverse structure (e.g.\ grain size) on a hundred nm length scale (as inferred from the width of the measured distribution). Already shortly after excitation changes of the scattering patterns can be recognized and three distinct features, in the following denoted as f1, f2, and f3 (marked in the frame for $\Delta t$ = 840\,ps), develop on different time scales. The most prominent transient feature - f1 - are the strong arc-shaped diffraction peaks at $q_v \approx \pm 21\,\mu$m$^{-1}$ in the vertical direction. They indicate the formation of a well-oriented structure (perpendicular to the laser polarization) with $\approx$ 300\,nm periodicity. Feature f2 corresponds to the (broader) diffraction peaks at $q_h \approx \pm 15 - 25\,\mu$m$^{-1}$ along the horizontal direction (at later times close to the beam-stop). Additionally, a broad, ring-like diffraction feature (f3) can be recognized. It has an elliptical shape with (at $\Delta t = 840$\,ps) vertical and horizontal half axes of $\approx$43\,$\mu$m$^{-1}$ and $\approx$36\,$\mu$m$^{-1}$, corresponding to real space periodicities of $\approx$146\,nm and $\approx$175\,nm, respectively.

\begin{figure}[htb]
\centering
\includegraphics[width=15cm]{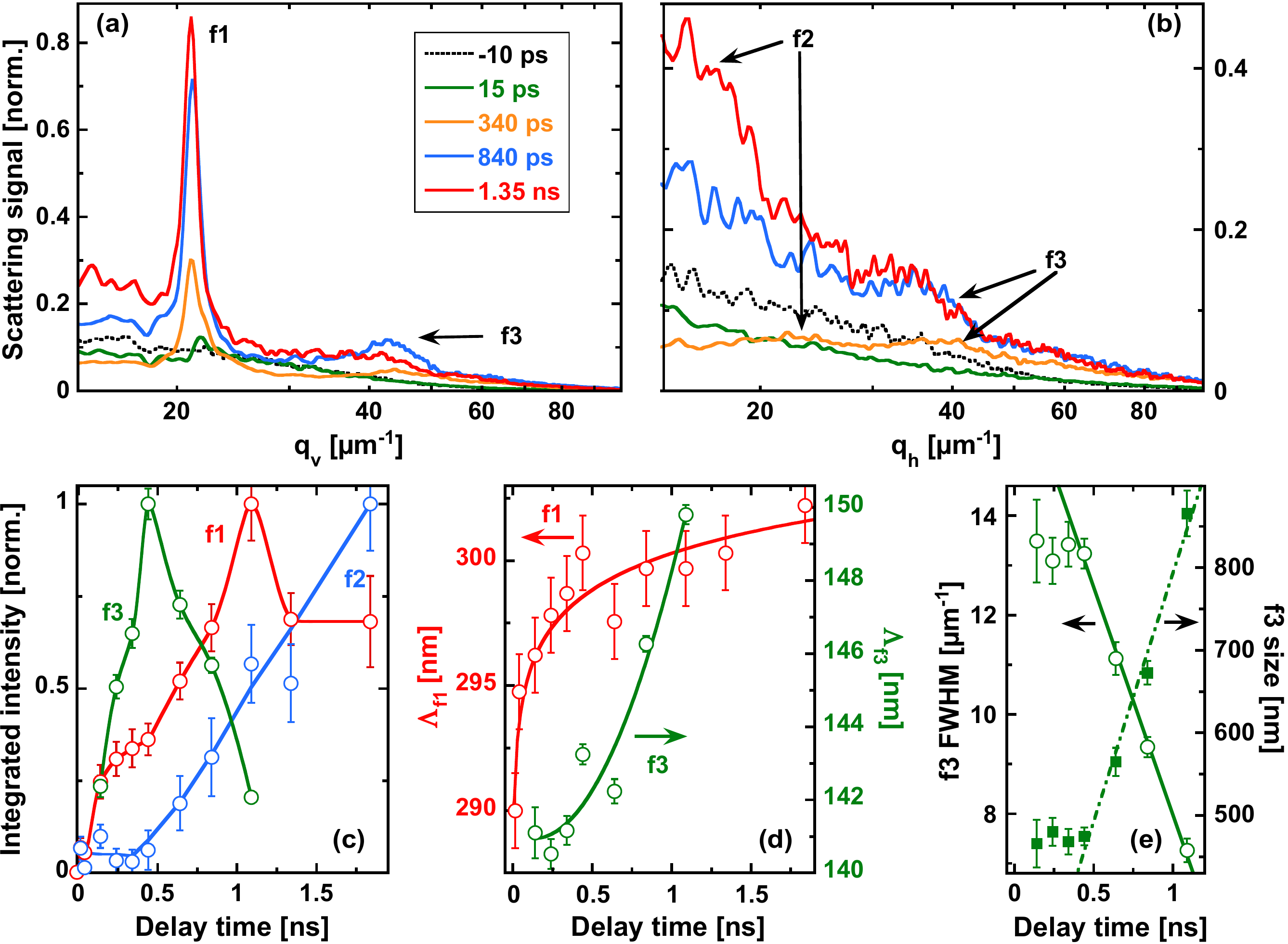}
\caption{\label{lipps_time} Top row: (a) Vertical ($I\left(q _v\right)$) and (b) horizontal ($I\left(q _h\right)$) cross sections along the white-dashed lines, as shown in the first frame of Fig.\ \ref{lipps_series}, for selected pump-probe delay times. Bottom row: (c) Integrated intensity of the characteristic features f1 (red), f2 (blue) and f3 (green), (d) real space periodicities $\rm \Lambda _{f1}$ of feature f1 (red, left axis) and $\rm \Lambda _{f3}$ of feature f3 (in vertical direction, green, right axis), (e) FWHM of feature f3 (open circles, left axis) and the derived size of the corresponding {\it structurally coherent} sample area (full squares, right axis), all as a function of pump-probe delay time. The solid curves in (c) and (d) are a guide to the eye; the curves in (e) represent linear fits to the data points at late delay times.}
\end{figure}

More detailed/quantitative information about the temporal evolution of the different diffraction features is presented in Fig.\ \ref{lipps_time}. The top row depicts vertical (a) and horizontal (b) cross sections along the white-dashed lines, as shown in the first frame of Fig.\ \ref{lipps_series}, for selected pump-probe time delays. The bottom row displays time-dependencies of (c) the integrated diffraction signal of f1 (red), f2 (blue), and f3 (green), of (d) the (vertical) real space periodicities $\rm \Lambda _{f1}$ (red) and $\rm \Lambda _{f3}$ (green) of features f1 and f3, respectively, and of (e) the peak width (FWHM) of f3 (open circles) along the vertical direction and, derived from this width the effective size of the corresponding {\it structurally coherent} sample area (filled squares; explanation see below).

Feature f1 starts to form already during the pump laser pulse (2$\rm ^{nd}$ frame in Fig.\ \ref{lipps_series}; green trace $I\left(q _v\right)$ in Fig.\ \ref{lipps_time}(a)). Its intensity increases with time (red curve in Fig.\ \ref{lipps_time}(c)) and it persists until the whole sample starts to disintegrate (last frame in Fig.\ \ref{lipps_series}, $\Delta t$ = 4.5\,ns). Moreover, there is a continuous shift in the peak position indicating an increase of its characteristic periodicity $\rm \Lambda _{f1}$ (red trace in Fig.\ \ref{lipps_time}(d)) with time. Feature f2 becomes visible after about 300\,ps and its intensity increases subsequently (green trace in Fig.\ \ref{lipps_time}(c)). Simultaneously f2 moves towards smaller $q$ but an accurate determination of the peak position as a function of delay time is difficult since at early delays it overlaps with feature f3, while at later delays it is partially blocked by the beam-stop. The elliptical ring feature f3 starts to develop after approximately 100\,ps, is most visible after a few hundred ps, and does not persist as long as features f1 and f2. Moreover, its peak position changes with time, being constant for about 400 ps before starting to shift to smaller $q_v$, which corresponds to an increasing real space periodicity $\rm \Lambda _{f3}$ (green trace in Fig.\ \ref{lipps_time}(d)). Also the width of the ring-like feature f3 changes with time, being constant again for approximately 400\,ps followed by a linear decrease, which could followed until f3 has almost vanished after about 1.4\,ns.

All quantities shown in Figs.\ \ref{lipps_time}(c) - (e) for the different features reveal essentially a two-stage evolution of the whole process. During the first stage up to about 0.4\,ns the intensity of f1 and f3 increases, while the intensity of f2 is low. During the same time $\rm \Lambda _{f1}$ increases, but $\rm \Lambda _{f3}$ stays constant. For delay times $\Delta t >$ 0.4\,ns (stage 2) f1 exhibits a further increase in intensity, which levels off around 1\,ns, while the intensity of f2 starts to continuously increase, and f3 decreases in intensity. Similarly, while $\rm \Lambda _{f1}$ remains essentially constant, $\rm \Lambda _{f3}$ increases and the f3 peak width decreases.

\section{Discussion and Interpretation of the transient Scattering Patterns}
Before discussing and interpreting the experimental results we would like to point out that {\it any} spatially dependent modulation of the optical properties (in the XUV!) of the laser-excited material will result in scattering of the probe beam. Such modulation can be caused by a variety of processes, such as electronic excitation, ionization, heating, but also changes in density and surface topography due to material expansion and ablation. Therefore, it is a priori difficult if not impossible to disentangle these processes from the scattering data without further knowledge, for example through their characteristic time-scales. Most likely the scattering observed at early delay times has to be attributed to electronic excitation and melting of the material, whereas at later times density modulations and changes of the surface topography due to expansion and ablation dominate.

\begin{figure}[htb]
\centering
\includegraphics[width=12cm]{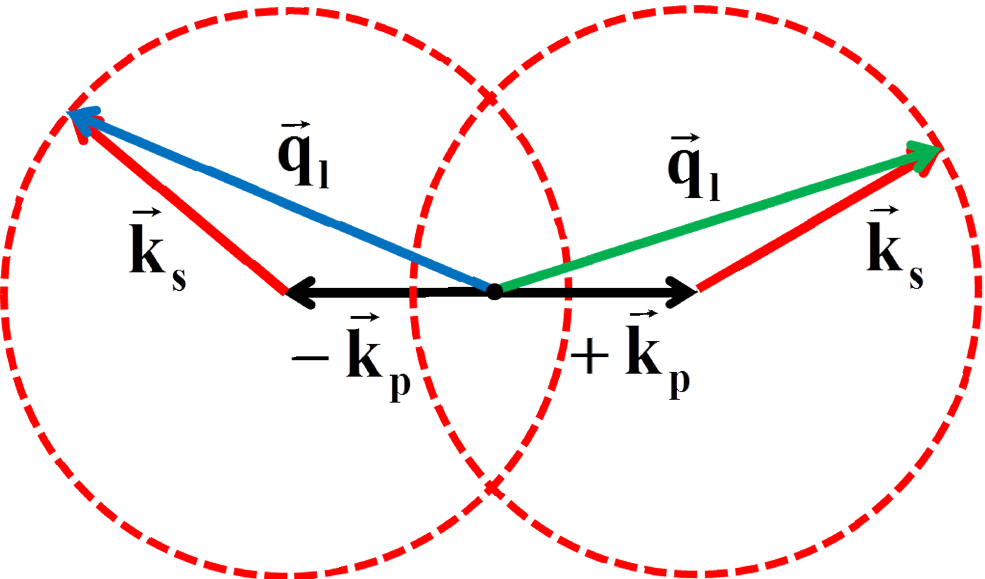}
\caption{\label{lipps_scheme} Interference model for LIPSS-generation: $\vec{k}_p$: in-surface component of the wave-vector $\vec{k}_i$ of the incident pump laser beam with $k_p = k_i\cdot \sin\left(\alpha \right)$; $\vec{k}_s$: wave-vector of a scattered/excited wave propagating along the surface; $\vec{q}_{l}$: LIPSS wave-vector.}
\end{figure}

To interpret the experimental data and the different diffraction features, we apply first the intuitive model of interference between the incident laser field and surface scattered/excited waves since it allows simple predictions on the periodicity and angular orientation of the resulting LIPSS. This is illustrated in Fig.\ \ref{lipps_scheme} where $\vec{k}_p$ denotes the in-surface component of the wave-vector $\vec{k}_i$ of the incident pump laser beam ($k_p = k_i\cdot \sin\left(\alpha\right)$; $\alpha$: angle of incidence), and $\vec{k}_s$ the wave-vector of a scattered/excited wave propagating along the surface.

While $\vec{k}_s$ can have, in principle, any in-plane direction, its length is determined by the particular nature of the scattering/excitation process. Often it has been found that $k_s \approx k_i$ and the LIPSS wave-vector $\vec{q}_{l}$ is then given by $\vec{q}_{l} = \pm \left(\vec{k}_p + \vec{k}_s\right)$. Therefore, all possible $\vec{q}_{l}$ lie on two intersecting circles with radius $k_p \approx k_i$ centered at $\pm \vec{k}_p$. Moreover, in many experiments (e.g.\ \cite{young83b}) three dominating LIPSS-modes were found for p-polarized light. The first two modes $q^{\pm}_{l}$ correspond to $\vec{k}_s$ parallel or anti-parallel to $\vec{k}_p$, the third mode $q^{0}_{l}$ to the intersection points of the two circles. For these cases $q^{\pm}_{l} = \left(k_i \pm k_p\right) = k_i\cdot\left( 1\pm \sin\left(\alpha \right)\right)$ and $q^{0}_{l} = k_i\cdot \cos\left(\alpha \right)$, which results in $q^{+}_{l} = 20.8\,\mu$m$^{-1}$, $q^{-}_{l} = 3.2\,\mu$m$^{-1}$, and $q^{0}_{l} = 8.1\,\mu$m$^{-1}$ here. The corresponding LIPSS-periodicities $\Lambda_l = 2\pi /q_l$ are $\Lambda ^{+}_l$ = 302\,nm, $\Lambda ^{-}_l$ = 1.95\,$\mu$m and $\Lambda ^{0}_l$ = 780\,nm. While $q^{-}_{l}$ and $q^{0}_{l}$ are blocked by the beam-stop, it seems natural to identify the arc-shaped diffraction feature f1 as segments of the intersecting circles and the f1-peak in the $I\left(q _v\right)$-profiles (Fig.\ \ref{lipps_time}(a)) as the $q^{+}_{l}$ LIPSS-mode.

\begin{figure}[htb]
\centering
\includegraphics[width=15cm]{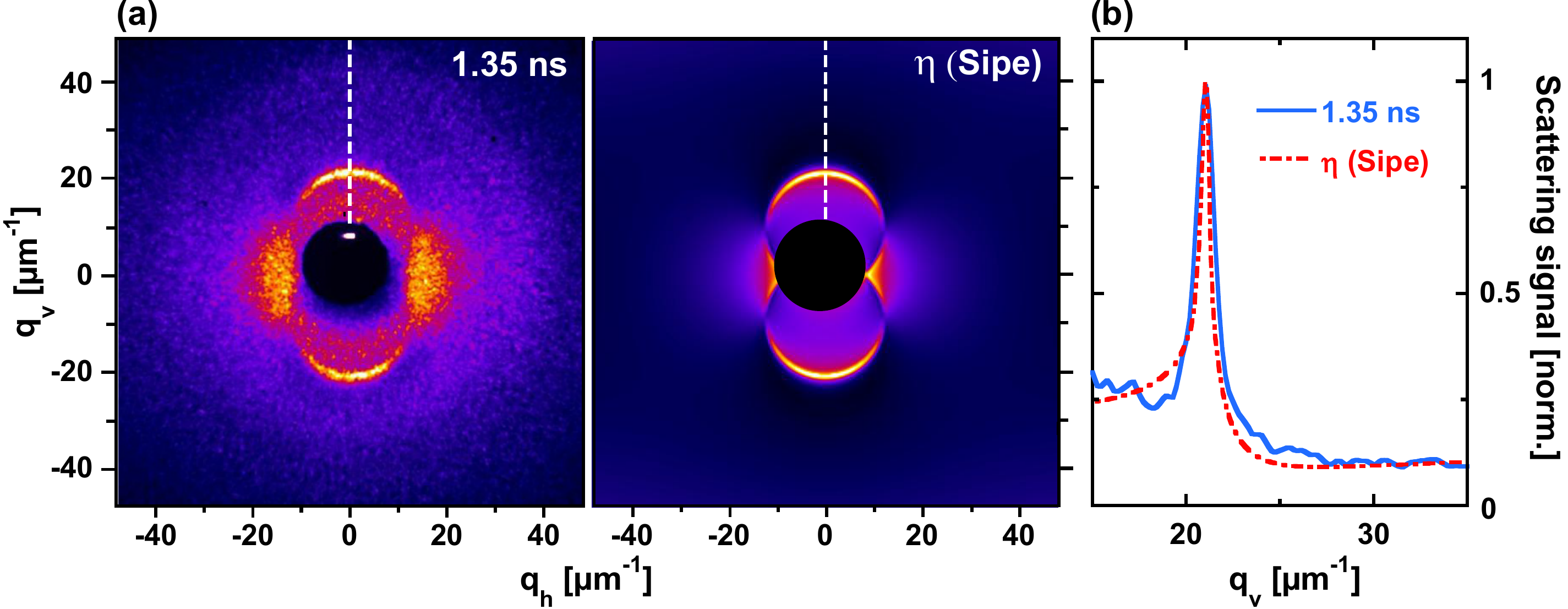}
\caption{\label{lipps_model} (a): Comparison of the transient scattering pattern measured at $\Delta t$ = 1.35\,ns and calculations of the so-called "efficacy factor" $\eta$ based on the model by Sipe et al.\ \cite{sipe83}. (b): Vertical cross sections (normalized) in the vicinity of the $q^{+}_{l}$-peak (red: $\eta \left( q_v\right)$; blue: $ I\left( q_v, 1.35\,{\rm ns}\right)$.}
\end{figure}

Going beyond the simplified interference-model the theory of Sipe et al.\ is applied \cite{sipe83, bonse05}. It allows to calculate the so-called {\it efficacy factor} $\rm \eta (\vec{q})$, which is proportional to the spatially varying deposition of the laser pulse energy (represented in Fourier space). Among other factors it depends on the optical constants of the irradiated material (for further details the reader is referred to \cite{bonse05, bonse09}). Figure\ \ref{lipps_model}(a) compares the transient scattering pattern measured at $\Delta t$ = 1.35\,ns to a calculated distribution of $\eta$  for the experimental parameters of our experiment (i.e.\ laser wavelength, polarization and angle of incidence) using the optical constants of liquid Si as input.

The agreement between the calculated efficacy factor and the measured scattering pattern is striking and almost quantitative (at least in certain {\it q}-ranges) as indicated by the vertical cross sections around the $q^{+}_{l}$-peak shown in Fig.\ \ref{lipps_model}(b). This unambiguously confirms the assignment of f1 made above. In the classification of the recent LIPSS literature \cite{bonse20b}, f1 is associated with type-s LSFL. It should be noted that the good agreement between the calculated $\eta$ and the measured experimental diffraction pattern is achieved only when using the optical constants of liquid Si (which is metallic) as input. Given the high pump fluence well above the melting and ablation threshold), this appears not surprising because melting will occur already during the 12\,ps duration laser pulse.

However, the efficacy factor $\rm \eta\left(\vec{q}\right)$ does not provide information on how and on which time scales such a modulated energy deposition transforms into modifications of the sample properties and morphology responsible for the transient scattering measured in the experiment. Moreover, the Sipe-model, as applied here, does not take into account changes of the sample (morphology, optical properties) during irradiation, which are very likely to occur for the conditions of our experiment (12\,ps pulse duration, high fluence) when the material transforms already during the pulse from a cold solid, over a hot, electronically excited semiconductor into a metallic liquid. Such changes will definitely influence the process of LIPSS-formation (e.g.\ Bonse et al.\ \cite{bonse09} have argued for the case of fs laser excited Si that changes of the optical properties caused by the formation of a dense electron-hole plasma can change the LIPSS-periodicity).

However, we believe that the increase of the real space periodicity $\rm \Lambda _{f1}$ of the f1-feature (Fig.\ \ref{lipps_time}(b)) observed here is only an apparent one since this would imply a corresponding expansion of the LIPSS-pattern over the whole pump laser spot size of $40\,\mu{\rm m}$ in horizontal direction, requiring lateral expansion velocities initially in excess of 5\,km/s, which appears unrealistic.

Pfau et al. \cite{pfau12} have observed changes of positions of diffraction peaks of similar magnitude (i.e.\ a few percent) in time-resolved magnetic scattering experiments (performed also at FLASH) during the ultrafast, laser-induced demagnetization of a thin-film Co/Pt multilayer. These samples exhibit a pronounced "worm-like" structure of domains with opposite magnetization, which gives rise to a characteristic diffraction ring. The peak position of this diffraction ring shifts to lower $q$ within a few hundred femtoseconds after laser excitation, which is attributed to changes of the scattering form factor due to a smoothing of the magnetization profile at the domain boundary caused by spin-dependent carrier transport.

We think, a similar form factor effect is responsible for the observed shift of the f1-peak - although on a very different time scale. The spatially periodic energy deposition (as described by the efficacy factor) transfers - very likely in a non-linear fashion - into a corresponding modulation of the material properties (e.g.\ temperature, melt depth) thus imposing lateral gradients of these properties. These will smooth out with time due to lateral transport (e.g.\ heat conduction), which may affect the form factor in a similar way as discussed in \cite{pfau12}.

Considering only Fig.\ \ref{lipps_model} it is tempting to interpret f2 also with the help of the Sipe-model and the efficacy factor as the "tail" of scattering signal originating from the vicinity of $\rm q^{0}_{l}$/$\rm q^{-}_{l}$. We exclude this explanation because the f2-feature appears significantly later than f1 and is initially well separated from the beam-stop (see the orange $I\left(q _h\right)$-trace in Fig.\ \ref{lipps_time}) and thus $\rm q^{0}_{l}$. However, the limited signal-to-noise ratio of the data in the {\it q}-range of the f2-feature does not allow for further quantitative analysis, so we refrain from speculating on its physical origin at this point.

Similarly, the calculated $\eta$ does not account for the elliptical f3-feature, which is strongest for $\Delta t$ < 1\,ns and which has almost vanished at $\Delta t$ = 1.35\,ns. As a possible origin of the f3 feature we suggest the formation of capillary waves at the surface of the laser-molten silicon film \cite{keilmann83, young84}, presumably excited through the recoil of ablating material. This scenario is supported by the timescales of the appearance of this feature at delay times $\Delta t \approx$ 100\,ps \cite{sokol98b}. For a quantitative analysis, we neglect gravitational forces (i.e.\ large capillary constant $a = \sqrt{\frac{2\sigma}{g\rho}}$ \cite{landau87}; $g$ gravitational acceleration, $\sigma$ surface tension, $\rho$ density) and apply the dispersion relation of capillary waves in the short wavelength limit (eq.\ 62.3 in \cite{landau87}). This results in an oscillation period $T = \sqrt{\frac{\Lambda ^3 \rho}{2\pi\sigma}}$ ($\Lambda$ wavelength of the capillary waves).

Identifying $\Lambda$ with $\Lambda _{f3} \approx$ 140\,nm at early delay times (stage 1) and using published values for the density $\rho = 2.52$\,g/cm$^3$ and the surface tension $\sigma = 0.874$ kg/s$^2$ of liquid Silicon at the melting point \cite{nakamura92} we obtain a half oscillation period of $T/2 \lesssim 0.6$\,ns, in reasonable agreement with the temporal evolution (rise and subsequent decay) of the intensity of the f3 feature \cite{keilmann83}. Within this picture f3 does not reappear since these short wavelength capillary waves are damped by viscous forces with a damping time $\tau _D = \frac{\rho \Lambda ^2}{8\pi ^2 \eta}\approx 2.8$\,ns ($\eta \approx 0.8$\,mPa$\cdot$s: dynamic viscosity of liquid Silicon \cite{kolasinski07}) and additionally laser-induced ablation decomposes the molten silicon film on the ns timescale. With respect to the observed time-dependent increase of $\Lambda _{f3}$ similar form factor changes might be responsible as discussed above for the LIPSS-periodicity $\Lambda _{f1}$.

Finally, we interpret the peak width $w$ (FWHM) of the f3 feature, as depicted by the open circles in Fig.\ \ref{lipps_time}(e), in a Scherrer-like approach \cite{scherrer18} as a measure of the size $s \approx 2\pi /w$ of the area, where capillary waves are {\it coherently} excited. This size (filled squares in Fig.\ \ref{lipps_time}(e)) has an approximately constant value of 470\,nm during stage 1, but increases linearly with a slope of $v = 600$\,m/s (dashed-dotted curve in Fig.\ \ref{lipps_time}(e)) during stage 2. Within the capillary wave picture $v$ corresponds to twice the group velocity $v_G$ of the excited waves assuming that the initial excitation area expands radially by capillary wave propagation. However, one must take into account that the value of $a$ (and thus $v$) derived from the f3 peak width $w$ with the equation above tends to overestimate the actual size for two reasons \cite{langford78}: (i) The Scherrer-equation includes a pre-factor $K$, which depends on the shape of the scattering object and is usually somewhat smaller than $1$; (ii) The determined value of $a$ represents in our case (2D scattering objects) an area-weighted mean value, which depends on the size distribution (unknown), but emphasizes larger {\it particles}. Therefore, $v/2 = 300$\,m/s represents actually an upper boundary for the group velocity $v_G$ of capillary waves, which can be theoretically calculated from the dispersion relation as $v_G = d\omega /dk = \frac{3}{2} \cdot \sqrt{\frac{2\pi \sigma}{\rho \Lambda}}$. Using again the published values for density $\rho$ and surface tension $\sigma$ of liquid Silicon \cite{nakamura92} we obtain $v_G\approx 180$\,m/s, in reasonable agreement with the upper boundary $v/2 = 300$\,m/s determined from the observed decrease of the f3 peak width.

In summary, although we ignore any temperature dependent changes of the thermophysical material properties the simple capillary wave picture provides a surprisingly good quantitative description of the temporal evolution of the f3 feature. However, two questions remain unanswered: (i) What mechanism selects the observed wavelength $\Lambda _{f3}$, and (ii), why is the f3 diffraction ring elliptical? As we will argue in the next chapter, both effects may be related to the f1 feature and the formation of type-s LSFL.

\section{Polarization Dependence}
\begin{figure}[htb]
\centering
\includegraphics[width=15cm]{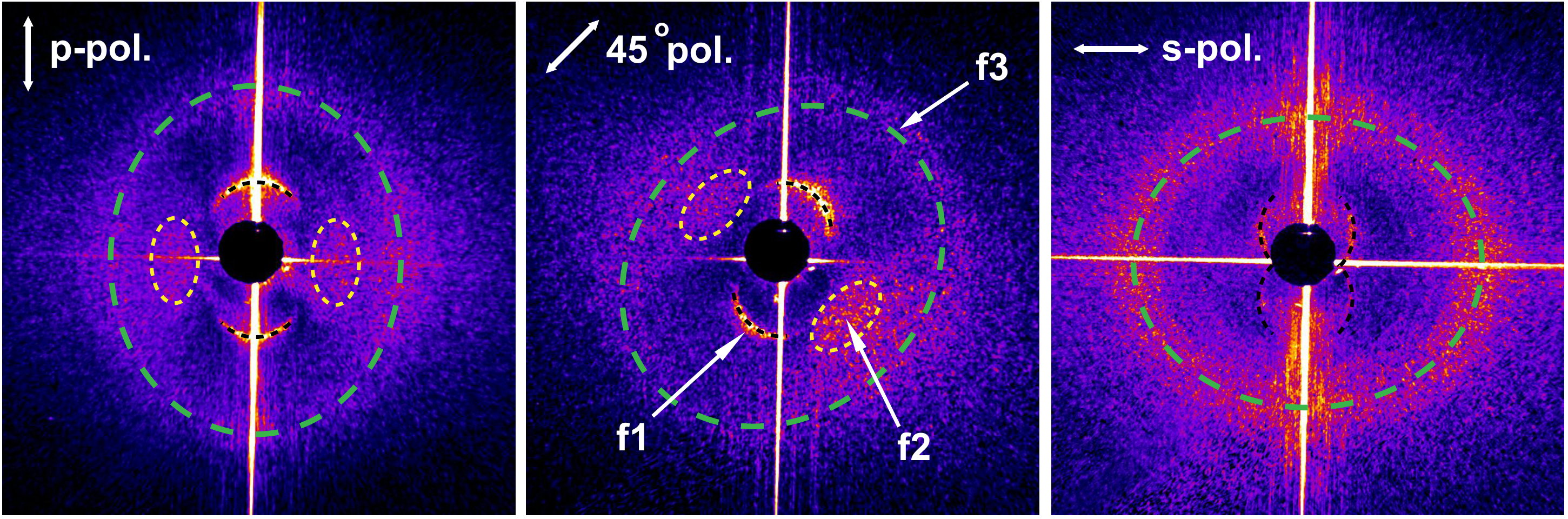}
\caption{\label{lipss_pol} Transient scattering patterns of a 100\,nm polycrystalline Si-film (on a $20 \times 20\,\mu$m$^2$, 20\,nm thick Si$_3$N$_4$-membrane) measured 340\,ps after excitation with a 12\,ps, 523\,nm laser pulse (fluence $\approx$ 2\,J/cm$^2$) for different polarization directions (left: p-pol., middle: 45$\rm ^o$-pol., right: s-pol.). The black, yellow and green dashed curves mark the characteristic diffraction features f1, f2 and f3, respectively. Please note, that the intensity scale has been adjusted for each image separately in order to make the different diffraction features clearly visible. The overall scattering intensity is decreasing by changing the pump polarization from p to s.}
\end{figure}

The Sipe-model allows predictions how the LIPSS-formation depends on laser-polarization \cite{sipe83,young83b}. In particular, it is expected (and has been experimentally confirmed) that upon rotation of the pump laser polarization the arc-segments, where $\rm \eta \left( \vec{q}\right)$ is maximum, rotate as well. To validate this, the XUV-scattering on a similar 100\,nm thick Si-film for similar excitation conditions \cite{note1} as above has been measured for three different polarization directions (p-, s- and 45$\rm ^o$-polarization). Data for a pump-probe delay of $\Delta t$ = 340\,ps are shown in Fig.\ \ref{lipss_pol}.

It has to be noted that these measurements have been performed with Silicon films deposited on smaller, $20 \times 20\,\mu$m$^2$-sized Si$_3$N$_4$-membranes. Diffraction from the edges of the membranes is responsible for the very intense vertical and horizontal streaks (white), which partially overlap with the different scattering features discussed above. However, in complete agreement with the Sipe-model, the f1-feature moves along the two intersecting circles depicted in Fig.\ \ref{lipps_scheme} and is weakest for s-polarized pump light. Although the f2-feature is not visible for s-polarized excitation, presumably because the scattering intensity is low (as for f1) and superimposed by the strong scattering from the membrane-edges, Fig.\ \ref{lipss_pol} clearly demonstrates that the diffraction features f2 and f3 change their orientation as well. However, while the motion of f1 is centered around $\rm \pm\vec{k}_p$, f2 and f3 rotate around the origin $\rm \vec{q} = 0$. As such, the elliptical f3 feature keeps its large axis always parallel to the linear laser beam polarization (the aspect ratio does not change), while the f2 feature is always oriented perpendicular to it.

To explain this behavior, one has to consider that the modulated energy deposition at the spatial frequency $q^{+}_{l}$ of the f1 feature, which leads to LIPSS-formation, also imposes an overall asymmetry of the transient material properties and consequently in the material dynamics between directions along and perpendicular to the pump laser polarization. Since the f1 feature appears first (already during the pulses), it may be not surprising that the processes leading to the f2 and f3 diffraction features depend on the pump pulse polarization in the observed manner (rotation with the polarization, ellipticity of f3), although f2 and f3 are not accounted for by the efficacy factor $\rm \eta\left(\vec{q}\right)$. In fact, in Chapter 5 of this book Rudenko and Colombier discuss various mechanisms (e.g.\ near field effects) that lead to a polarization dependent symmetry breaking and thus preferred orientation of the evolving structures. In addition, it is noteworthy that $\Lambda _{f3} \approx \Lambda _{f1}/2$, which points towards a further connection between the f1 and f3 features, suggesting that the so far not understood selection of wavelength within the capillary wave picture is also related to the periodically modulated energy deposition.

\section{Higher Order LIPSS}
\begin{figure}[htb]
\centering
\includegraphics[width=15cm]{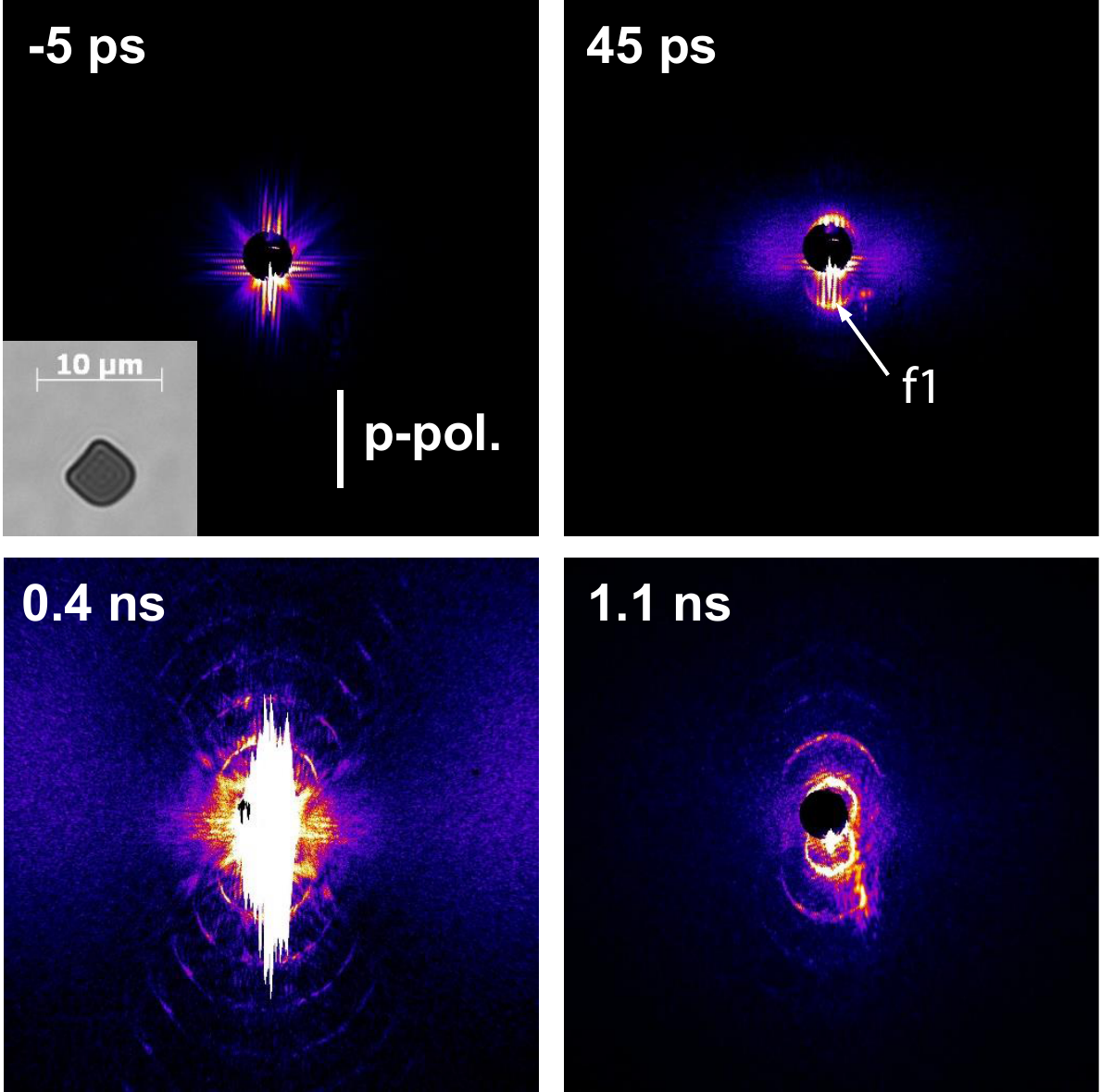}
\caption{\label{lipss_fs} Transient scattering pattern of $5\,\times\,5\,\mu$m$^2$ sized Si-patches of 100\,nm thickness deposited on free-standing 20\,nm thick Si$_3$N$_4$ membranes after excitation with a 50\,fs, 800\,nm laser pulse for different pump-probe delay times; solid white bar in the first frame: direction of laser polarization (projection onto the surface). The inset in the upper-left frame shows an optical micrograph of a typical sample.}
\end{figure}

In this section we discuss results from a scattering experiment in a similar experimental geometry, but using single 50\,fs, 800\,nm, and p-polarized laser pulses at an angle of incidence of 45$\rm ^o$ (p-pol.) for the excitation of approximately $5\,\times\,5\,\mu$m$^2$ sized Si-patches of 100\,nm thickness deposited on free-standing 20\,nm thick Si$_3$N$_4$ membranes (see inset in the upper-left frame in in Fig.\ \ref{lipss_fs}) at a fluence again well above the ablation threshold. For several reasons not be discussed here we were not able to perform systematic measurements, but only to record a limited number of scattering patterns. A short sequence of those are shown in Fig.\ \ref{lipss_fs} for different pump-probe time delays. As in Fig.\ \ref{lipps_series}, the projection of the pump laser beam polarization (p-pol.) onto the sample surface is indicated by the vertical solid white bar in the first frame of Fig.\ \ref{lipss_fs}.

Please note that in order to make diffraction features at larger scattering angles visible (e.g.\ scattering pattern at $\Delta t$ = 0.4\,ns) high FEL-intensities were required leading to saturation of the detector closer to the beam-stop. The saturated region was smeared out further in vertical direction due to the read-out process of the CCD-detector.

The upper-left pattern measured at negative time delay shows the scattering of the unperturbed Si-patch. Two features of the scattering patterns measured at positive delay times are particularly noteworthy:

(1) The f1-feature, which is characteristic for type-s LSFL as predicted by the Sipe-model, is observed even after excitation with a single fs pulse. Since the pump spot size on the sample is larger than the actual sample size, we attribute this to efficient scattering of the pump radiation at the edges of the Si-patch, similar to the observations on surfaces, which have been intentionally pre-structured/scratched \cite{kafka15, liu18}.

(2) As is best visible in the two scattering patterns at larger delays, not only the fundamental of the f1-feature is observed, but also higher diffraction orders (we were able to detect scattering up to the 9$^{\rm th}$ order) indicating that the spatial modulation of the material properties that give rise to the scattering of the XUV probe pulse is non-sinusoidal. These higher orders are observed at longer delay times and can be, therefore, very likely associated with changes of the sample morphology due to a spatially modulated ablation of the material. Since ablation exhibits generally a non-linear dependence on fluence (e.g.\ threshold behavior) a non-sinusoidal surface morphology will develop even for a sinusoidally modulated energy deposition leading to higher order diffraction.

We would like to emphasize that these results, although not yet systematic, specifically highlight the advantage of probing the LIPSS dynamics with short wavelength radiation. Higher order transient diffraction from LIPSS have so far been reported only in the work by Keilmann \cite{keilmann83}, who observed 2nd order LIPSS diffraction after excitation with a CO$_2$-laser (wavelength 10.6\,$\mu$m). In all other studies the accessible momentum transfer range was not sufficient to capture these details.

\section{Summary and Outlook}
The formation of LIPSS (and structure formation in general) at laser-irradiated surfaces represents a complex multiscale problem in terms of the relevant length and time scales, as well as the underlying physical processes, posing extreme challenges for their experimental and theoretical investigation.

On the experimental side techniques are required that combine high temporal resolution down to the femtosecond range with the appropriate spatial resolution on the nanometer to micrometer scale. Furthermore, these techniques need to be applied under highly irreversible conditions where single pulse laser excitation leads to significant permanent changes of the irradiated surface area.

In this contribution we have presented a set of time-resolved scattering experiments using femtosecond XUV-pulses from a FEL that allowed us to follow directly structure formation at the surface of thin Silicon films after irradiation with picosecond optical laser pulses on sub-micrometer to nanometer length scales.

Although the measured data reveal a complex evolution both, in time and momentum space, the main aspects are accounted for by the LIPSS-model of Sipe et al.\ \cite{sipe83}, which is confirmed almost quantitatively. However, the additional diffraction features observed here and their complex temporal evolution reveal the relevance of other processes (e.g.\ excitation of capillary waves).

Due to time-constraints - the presented data represent results from approx.\ 48 hours of FEL beamtime - more detailed and systematic studies had not been possible. Nevertheless, our experiments represent the first step towards more thorough investigations of nanoscale structure formation (incl.\ LIPSS) at laser-irradiated surface on extreme time and length scales that have not been accessible before.

While the extension of these experiments towards shorter probing wavelengths will enable measurements with even higher spatial resolution, the application of a grazing-incidence scattering geometry \cite{randolph20} will allow to study the surfaces of bulk materials, a situation more typical/relevant in {\it real life} laser processing. Similarly, the technique can be also applied in the multi-pulse excitation regime by pre-structuring the surface through irradiation with $N$ laser pulses and probing the transient changes induced by the $\left(N+1\right)$ excitation pulse, as has been done in all-optical pump-probe experiments (e.g.\ \cite{hoehm13, jia14}). On the other hand 3D-tomography using multiple probe pulses at different angles of incidence may provide volumetric information.

Moreover, experimental schemes that make use of the very high spatial coherence of the FEL-radiation, like coherent diffraction imaging \cite{chapman06} and photon correlation spectroscopy (e.g.\ \cite{roseker18, sun21}) can provide additional opportunities. In fact, during the same beamtime, where the data presented here have been measured, we have additionally used both techniques to follow in real space with nanometer resolution and on picosecond time scales the laser-induced destruction of a fabricated nanostructure \cite{barty08}.

With respect to the theoretical description of these processes such experiments, in particular if scattering intensities can be properly calibrated/normalized, should provide important benchmark data for the sophisticated hybrid electromagnetic and hydrodynamic continuum models (see for example \cite{rudenko20} and Chapter 5 of this book) that have been developed recently.

\acknowledgements{We are deeply indebted to Bruce Woods for his inventive engineering solutions that were critical for the success of these experiments, and his skill and efficiency in managing the construction of the instrumentation. We dedicate this to his memory. We also thank H. Ehrke, M. M. Seibert, and R. Tobey for their help during the experiment. This work was financially supported by the following agencies: The {\it Deutsche Forschungsgemeinschaft} (DFG, German Research Foundation) through the Collaborative Research Centre (CRC) 1242 (project number 278162697, project C01 {\it Structural Dynamics in Impulsively Excited Nanostructures}); The German Federal Ministry of Education and Research (FSP 301, grant 05KS7PG1); The US Department of Energy (DOE) Lawrence Livermore National Laboratory; The National Science Foundation Center for Biophotonics, University of California, Davis; The Advanced Light Source and National Centre for Electron Microscopy, Lawrence Berkeley Laboratory, under contract DE-AC03-76SF00098; The Natural Sciences and Engineering Research Council of Canada (NSERC Postdoctoral Fellowship to M.J.B.); The Sven and Lilly Lawskis Foundation (doctoral fellowship to M.M.S.); The US Department of Energy Office of Science to the Stanford Linear Accelerator Center; The European Development Fund (Structural dynamics of biomolecular systems - ELIBIO, CZ.02.1.01/0.0/0.0/15\_003/0000447); The European Union (TUIXS); The Swedish Research Council; The Swedish Foundation for International Cooperation in Research and Higher Education; The Swedish Foundation for Strategic Research. This work was performed under the auspices of the US DOE by Lawrence Livermore National Laboratory in part under contract W-7405-Eng-48 and in part under contract DE-AC52-07NA27344.}
%



\begin{thebibliography}{10}

\bibitem{rethfeld04}
B.~Rethfeld, K.~Sokolowski-Tinten, D.~von~der Linde, S.I. Anisimov, Appl. Phys.
  A: Mater. Sci. Process. \textbf{79}, 767 (2004).
\newblock \doi{10.1007/s00339-004-2805-9}

\bibitem{birnbaum65}
M.~Birnbaum, J. Appl. Phys. \textbf{36}(11), 3688 (1965).
\newblock \doi{10.1063/1.1703071}

\bibitem{florian20}
C.~Florian, S.V. Kirner, J.~Kr\"uger, J.~Bonse, J. Laser. Appl. \textbf{32}(2),
  022063 (2020).
\newblock \doi{10.2351/7.0000103}

\bibitem{emmony73}
D.C. Emmony, R.P. Howson, W.L. J., Applied Physics Letters \textbf{23}(11), 598
  (1973).
\newblock \doi{10.1063/1.1654761}

\bibitem{sipe83}
J.E. Sipe, J.F. Young, J.S. Preston, H.M. van Driel, Phys. Rev. B \textbf{27},
  1141 (1983).
\newblock \doi{10.1103/PhysRevB.27.1141}

\bibitem{dufft09}
D.~Dufft, A.~Rosenfeld, S.K. Das, R.~Grunwald, J.~Bonse, J. Appl. Phys.
  \textbf{105}(3), 034908 (2009).
\newblock \doi{10.1063/1.3074106}

\bibitem{skolski10}
J.Z.P. Skolski, G.R.B.E. R\"omer, J.V. Obona, V.~Ocelik, A.J. Huis in~'t Veld,
  J.T.M. De~Hosson, Phys. Rev. B \textbf{85}, 075320 (2012).
\newblock \doi{10.1103/PhysRevB.85.075320}

\bibitem{hoehm12}
S.~H\"{o}hm, A.~Rosenfeld, J.~Kr\"{u}ger, J.~Bonse, J. Appl. Phys.
  \textbf{112}, 014901 (2012).
\newblock \doi{10.1063/1.4730902}

\bibitem{rudenko17}
A.~Rudenko, J.P. Colombier, S.~H\"ohm, A.~Rosenfeld, J.~Kr\"uger, J.~Bonse,
  T.E. Itina, Sci. Rep. \textbf{7}, 12306 (2017).
\newblock \doi{10.1038/s41598-017-12502-4}

\bibitem{jia05}
T.Q. Jia, H.X. Chen, M.~Huang, F.L. Zhao, J.R. Qiu, R.X. Li, Z.Z. Xu, X.K. He,
  J.~Zhang, H.~Kuroda, Phys. Rev. B \textbf{72}, 125429 (2005).
\newblock \doi{10.1103/PhysRevB.72.125429}

\bibitem{guo08}
X.~Guo, R.~Li, Y.~Hang, Z.~Xu, B.~Yu, H.~Ma, B.~Lu, X.~Sun, Mater. Lett.
  \textbf{62}, 1769 (2008).
\newblock \doi{10.1016/j.matlet.2007.09.082}

\bibitem{costache04}
F.~Costache, S.~Kouteva-Arguirova, J.~Reif, Appl. Phys. A: Mater. Sci. Process.
  \textbf{79}, 1429 (2004).
\newblock \doi{10.1007/s00339-004-2803-y}

\bibitem{reif08}
J.~Reif, O.~Varlamova, F.~Costache, Appl. Phys. A: Mater. Sci. Process.
  \textbf{92}, 1019 (2008).
\newblock \doi{10.1007/s00339-008-4671-3}

\bibitem{rudenko20}
A.~Rudenko, A.~Abou-Saleh, F.~Pigeon, C.~Mauclair, F.~Garrelie, R.~Stoian, J.P.
  Colombier, Acta Mater. \textbf{194}, 93 (2020).
\newblock \doi{10.1016/j.actamat.2020.04.058}

\bibitem{bonse17}
J.~Bonse, S.~H\"ohm, S.V. Kirner, A.~Rosenfeld, J.~Kr\"uger, IEEE J. Sel. Top.
  Quantum Electron. \textbf{23}, 9000615 (2017).
\newblock \doi{10.1109/JSTQE.2016.2614183}

\bibitem{bonse20}
J.~Bonse, Nanomaterials \textbf{10}, 1950 (2020).
\newblock \doi{10.3390/nano10101950}

\bibitem{bonse20b}
J.~Bonse, S.~Gr\"af, Laser \& Photonics Reviews \textbf{14}, 2000215 (2020).
\newblock \doi{10.1002/lpor.202000215}

\bibitem{ehrlich82}
D.J. Ehrlich, S.R.J. Brueck, J.Y. Tsao, Appl. Phys. Lett. \textbf{41}, 630
  (1982).
\newblock \doi{10.1063/1.93631}

\bibitem{keilmann83}
F.~Keilmann, Phys. Rev. Lett. \textbf{51}, 2097 (1983).
\newblock \doi{10.1103/PhysRevLett.51.2097}

\bibitem{young83a}
J.F. Young, J.S. Preston, J.E. Sipe, H.M. van Driel, Phys. Rev. B \textbf{27},
  1424 (1983).
\newblock \doi{10.1103/PhysRevB.27.1424}

\bibitem{young84}
J.F. Young, J.E. Sipe, H.M. van Driel, Phys. Rev. B \textbf{30}, 2001 (1984).
\newblock \doi{10.1103/PhysRevB.30.2001}

\bibitem{lee92}
T.D. Lee, H.W. Lee, C.H. Nam, J.K. Kim, C.O. Park, J. Appl. Phys. \textbf{71},
  4208 (1992).
\newblock \doi{10.1063/1.350799}

\bibitem{hoehm13}
S.~H\"ohm, A.~Rosenfeld, J.~Kr\"uger, J.~Bonse, Appl. Phys. Lett. \textbf{102},
  054102 (2013).
\newblock \doi{10.1063/1.4790284}

\bibitem{murphy13}
R.D. Murphy, B.~Torralva, D.P. Adams, S.M. Yalisove, Appl. Phys. Lett.
  \textbf{103}, 141104 (2013).
\newblock \doi{10.1063/1.4823588}

\bibitem{jia14}
X.~Jia, T.Q. Jia, N.N. Peng, D.H. Feng, S.A. Zhang, Z.R. Sun, J. Appl. Phys.
  \textbf{115}, 143102 (2014).
\newblock \doi{10.1063/1.4870445}

\bibitem{kafka15}
K.R.P. Kafka, D.R. Austin, H.~Li, A.Y. Yi, J.~Cheng, E.A. Chowdhury, Opt.
  Express \textbf{23}, 19432 (2015).
\newblock \doi{10.1364/OE.23.019432}

\bibitem{garcia16}
M.~Garcia-Lechuga, D.~Puerto, Y.~Fuentes-Edfuf, J.~Solis, J.~Siegel, ACS
  Photonics \textbf{3}, 1961 (2016).
\newblock \doi{10.1021/acsphotonics.6b00514}

\bibitem{zhou17}
K.~Zhou, X.~Jia, T.~Jia, K.~Cheng, K.~Cao, S.~Zhang, D.~Feng, Z.~Sun, J. Appl.
  Phys. \textbf{121}, 104301 (2017).
\newblock \doi{10.1063/1.4978375}

\bibitem{liu18}
J.~Liu, X.~Jia, W.~Wu, K.~Cheng, D.~Feng, S.~Zhang, Z.~Sun, T.~Jia, Opt.
  Express \textbf{26}, 6302 (2018).
\newblock \doi{10.1364/OE.26.006302}

\bibitem{ayvazyan06}
V.~Ayvazyan, N.~Baboi, J.~B\"ahr, V.~Balandin, B.~Beutner, A.~Brandt,
  I.~Bohnet, A.~Bolzmann, R.~Brinkmann, O.~Brovko, J.~Carneiro, S.~Casalbuoni,
  M.~Castellano, P.~Castro, L.~Catani, E.~Chiadroni, S.~Choroba, A.~Cianchi,
  H.~Delsim-Hashemi, G.D. Pirro, M.~Dohlus, S.~D\"usterer, H.~Edwards,
  B.~Faatz, A.~Fateev, J.~Feldhaus, K.~Fl\"ottmann, J.~Frisch, L.~Fr\"ohlich,
  T.~Garvey, U.~Gensch, N.~Golubeva, H.J. Grabosch, B.~Grigoryan, O.~Grimm,
  U.~Hahn, J.~Han, M.~v.~Hartrott, K.~Honkavaara, M.~H\"uning, R.~Ischebeck,
  E.~Jaeschke, M.~Jablonka, R.~Kammering, V.~Katalev, B.K.S. Khodyachykh,
  Y.~Kim, V.~Kocharyan, M.~K\"orfer, M.~Kollewe, D.~Kostin, D.~Kr\"amer,
  M.~Krassilnikov, G.~Kube, L.~Lilje, T.~Limberg, D.~Lipka, F.~L\"ohl,
  M.~Luong, C.~Magne, J.~Menzel, P.~Michelato, V.~Miltchev, M.~Minty,
  W.~M\"oller, L.~Monaco, W.~M\"uller, M.~Nagl, O.~Napoly, P.~Nicolosi,
  D.~N\"olle, T.N. {n}ez, A.~Oppelt, C.~Pagani, R.~Paparella, B.~Petersen,
  B.~Petrosyan, J.~Pfl\"uger, P.~Piot, E.~Pl\"onjes, L.~Poletto, D.~Proch,
  D.~Pugachov, K.~Rehlich, D.~Richter, S.~Riemann, M.~Ross, J.~Rossbach,
  M.~Sachwitz, E.L.Saldin, W.~Sandner, H.~Schlarb, B.~Schmidt, M.~Schmitz,
  P.~Schm\"user, J.~Schneider, E.~Schneidmiller, H.J. Schreiber, S.~Schreiber,
  A.~Shabunov, D.~Sertore, S.~Setzer, S.~Simrock, E.~Sombrowski, L.~Staykov,
  B.~Steffen, F.~Stephan, F.~Stulle, K.~Sytchev, H.~Thom, K.~Tiedtke,
  M.~Tischer, R.~Treusch, D.~Trines, I.~Tsakov, A.~Vardanyan, R.~Wanzenberg,
  T.~Weiland, H.~Weise, M.~Wendt, I.~Will, A.~Winter, K.~Wittenburg, M.~Yurkov,
  I.~Zagorodnov, P.~Zambolin, K.~Zapfe, Eur. Phys. J. D \textbf{37}, 297
  (2006).
\newblock \doi{10.1140/epjd/e2005-00308-1}

\bibitem{note1}
Some of these results have been presented earlier in \cite{sokol10b}

\bibitem{sokol10b}
K.~Sokolowski-Tinten, A.~Barty, S.~Boutet, U.~Shymanovich, H.~Chapman,
  M.~Bogan, S.~Marchesini, S.~Hau-Riege, N.~Stojanovic, J.~Bonse, Y.~Rosandi,
  H.M. Urbassek, R.~Tobey, H.~Ehrke, A.~Cavalleri, S.~D\"usterer, H.~Redlin,
  M.~Frank, S.~Bajt, J.~Schulz, M.~Seibert, J.~Hajdu, R.~Treusch, C.~Bostedt,
  M.~Hoener, T.~M\"oller, AIP Conf. Proc. \textbf{1278}, 373 (2010).
\newblock \doi{10.1063/1.3507123}

\bibitem{bostedt09}
C.~Bostedt, H.N. Chapman, J.T. Costello, J.R. {Crespo López-Urrutia},
  S.~Düsterer, S.W. Epp, J.~Feldhaus, A.~Föhlisch, M.~Meyer, T.~Möller,
  R.~Moshammer, M.~Richter, K.~Sokolowski-Tinten, A.~Sorokin, K.~Tiedtke,
  J.~Ullrich, W.~Wurth, Nucl. Instrum. Methods Phys. Res. A \textbf{601}, 108
  (2009).
\newblock \doi{j.nima.2008.12.202}.
\newblock Special issue in honour of Prof. Kai Siegbahn

\bibitem{rossbach19}
J.~Rossbach, J.R. Schneider, W.~Wurth, Phys. Rep. \textbf{808}, 1 (2019).
\newblock \doi{j.physrep.2019.02.002}

\bibitem{tiedtke09}
K.~Tiedtke, A.~Azima, N.~von Bargen, L.~Bittner, S.~Bonfigt, S.~D\"usterer,
  B.~Faatz, U.~Fr\"uhling, M.~Gensch, C.~Gerth, N.~Guerassimova, U.~Hahn,
  T.~Hans, M.~Hesse, K.~Honkavaar, U.~Jastrow, P.~Juranic, S.~Kapitzki,
  B.~Keitel, T.~Kracht, M.~Kuhlmann, W.B. Li, M.~Martins, T.~N{\'{u}}{\~{n}}ez,
  E.~Pl\"onjes, H.~Redlin, E.L. Saldin, E.A. Schneidmiller, J.R. Schneider,
  S.~Schreiber, N.~Stojanovic, F.~Tavella, S.~Toleikis, R.~Treusch, H.~Weigelt,
  M.~Wellh\"ofer, H.~Wabnitz, M.V. Yurkov, J.~Feldhaus, New J. Phys.
  \textbf{11}, 023029 (2009).
\newblock \doi{10.1088/1367-2630/11/2/023029}

\bibitem{barty08}
A.~Barty, S.~Boutet, M.J. Bogan, S.~Marchesini, K.~Sokolowski-Tinten,
  N.~Stojanovic, R.~Tobey, H.~Ehrke, A.~Cavalleri, S.~D\"usterer, M.~Frank,
  S.~Bajt, B.W. Woods, M.M. Seibert, J.~Hajdu, R.~Treusch, , H.N. Chapman,
  Nature Photon. \textbf{2}, 415 (2008).
\newblock \doi{10.1038/nphoton.2008.128}

\bibitem{bajt08}
S.~Bajt, H.N. Chapman, E.A. Spiller, J.B. Alameda, B.W. Woods, M.~Frank, M.J.
  Bogan, A.~Barty, S.~Boutet, S.~Marchesini, S.P. Hau-Riege, J.~Hajdu,
  D.~Shapiro, Appl. Opt. \textbf{47}, 1673 (2008).
\newblock \doi{10.1364/AO.47.001673}

\bibitem{chapman06}
H.N. Chapman, A.~Barty, M.J. Bogan, S.~Boutet, M.~Frank, S.P. Hau-Riege,
  S.~Marchesini, B.W. Woods, S.~Bajt, W.H. Benner, R.A. London, E.~Pl\"onjes,
  M.~Kuhlmann, R.~Treusch, S.~D\"usterer, T.~Tschentscher, J.R. Schneider,
  E.~Spiller, T.~M\"oller, C.~Bostedt, M.~Hoener, D.A. Shapiro, K.O. Hodgson,
  D.~van~der Spoel, F.~Burmeister, M.~Bergh, C.~Caleman, G.~Huldt, M.M.
  Seibert, F.R. Maia, R.W. Lee, A.~Szoke, N.~Timneanu, J.~Hajdu, Nature Phys.
  \textbf{2}, 839 (2006).
\newblock \doi{10.1038/nphys461}

\bibitem{redlin11}
H.~Redlin, A.~Al-Shemmary, A.~Azima, N.~Stojanovic, F.~Tavella, I.~Will,
  S.~Düsterer, Nucl. Instr. Meth. Phys. Res. A \textbf{635}, S88 (2011).
\newblock \doi{10.1016/j.nima.2010.09.159}

\bibitem{singer08}
A.~Singer, I.A. Vartanyants, M.~Kuhlmann, S.~Duesterer, R.~Treusch,
  J.~Feldhaus, Phys. Rev. Lett. \textbf{101}, 254801 (2008).
\newblock \doi{10.1103/PhysRevLett.101.254801}

\bibitem{young83b}
J.F. Young, J.S. Preston, H.M. van Driel, J.E. Sipe, Phys. Rev. B \textbf{27},
  1155 (1983).
\newblock \doi{10.1103/PhysRevB.27.1155}

\bibitem{bonse05}
J.~Bonse, M.~Munz, H.~Sturm, J. Appl. Phys. \textbf{97}, 013538 (2005).
\newblock \doi{10.1063/1.1827919}

\bibitem{bonse09}
J.~Bonse, A.~Rosenfeld, J.~Kr\"uger, J. Appl. Phys. \textbf{106}, 104910
  (2009).
\newblock \doi{10.1063/1.3261734}

\bibitem{pfau12}
B.~Pfau, S.~Schaffert, L.~Müller, C.~Gutt, A.~Al-Shemmary, F.~B\"uttner,
  R.~Delaunay, S.~D\"usterer, S.~Flewett, R.~Fr\"omter, J.~Geilhufe, E.~Guehrs,
  C.M. Günther, R.~Hawaldar, M.~Hille, N.~Jaouen, A.~Kobs, K.~Li, J.~Mohanty,
  H.~Redlin, W.F. Schlotter, D.~Stickler, R.~Treusch, B.~Vodungbo, M.~Kl\"aui,
  H.P. Oepen, J.~L\"uning, G.~Gr\"ubel, S.~Eisebitt, Nat. Commun. \textbf{3},
  1100 (2012).
\newblock \doi{10.1038/ncomms2108}

\bibitem{sokol98b}
K.~Sokolowski-Tinten, J.~Bialkowski, A.~Cavalleri, D.~{von der Linde}, Applied
  Surface Science \textbf{127-129}, 755 (1998).
\newblock \doi{10.1016/S0169-4332(97)00736-8}

\bibitem{landau87}
L.D. Landau, E.M. Lifshitz, \emph{{Fluid Mechanics}}, \emph{Course of
  Theoretical Physics}, vol.~6, 2nd edn. (Pergamon, 1987)

\bibitem{nakamura92}
S.~Nakamura, T.~Hibiya, Int. J. Thermophys. \textbf{13}, 1061 (1992).
\newblock \doi{10.1007/BF01141216}

\bibitem{kolasinski07}
K.W. Kolasinski, Curr. Opin. Solid State Mater. Sci. \textbf{11}, 76 (2007).
\newblock \doi{10.1016/j.cossms.2008.06.004}

\bibitem{scherrer18}
P.~Scherrer, Nachrichten Gesellschaft der Wissenschaften zu G\"ottingen,
  Mathematisch-Physikalische Klasse pp. 98--100 (1918)

\bibitem{langford78}
J.I. Langford, A.J.C. Wilson, J. Appl. Crystallogr. \textbf{11}, 102 (1978).
\newblock \doi{10.1107/S0021889878012844}

\bibitem{note2}
For the measurements with 45$\rm^o$- and s-polarized pumping we increased the laser fluence to the maximum available level of about 2.2\,J/cm$^2$ to account at least partly for the increased reflection losses compared to pumping with p-polarized light.

\bibitem{randolph20}
L.~Randolph, M.~Banjafar, T.R. Preston, T.~Yabuuchi, M.~Makita, N.P. Dover,
  C.~Rödel, S.~Göde, Y.~Inubushi, G.~Jakob, J.~Kaa, A.~Kon, J.K. Koga,
  D.~Ksenzov, T.~Matsuoka, M.~Nishiuchi, M.~Paulus, F.~Schon, K.~Sueda,
  Y.~Sentoku, T.~Togashi, M.~Vafaee-Khanjani, M.~Bussmann, T.E. Cowan,
  M.~Kläui, C.~Fortmann-Grote, A.P. Mancuso, T.~Kluge, C.~Gutt,
  M.~Nakatsutsumi,   (2020).
\newblock ArXiv 2012.15076 (2020)

\bibitem{roseker18}
W.~Roseker, S.O. Hruszkewycz, F.~Lehmkühler, M.~Walther,
  H.~Schulte-Schrepping, S.~Lee, T.~Osaka, L.~Strüder, R.~Hartmann,
  M.~Sikorski, S.~Song, A.~Robert, P.H. Fuoss, M.~Sutton, G.B. Stephenson,
  G.~Grübel, Nat. Commun. \textbf{9}, 1704 (2018).
\newblock \doi{10.1038/s41467-018-04178-9}

\bibitem{sun21}
Y.~Sun, G.~Carini, M.~Chollet, F.J. Decker, M.~Dunne, P.~Fuoss, S.O.
  Hruszkewycz, T.J. Lane, K.~Nakahara, S.~Nelson, A.~Robert, T.~Sato, S.~Song,
  G.B. Stephenson, M.~Sutton, T.B. Van~Driel, C.~Weninger, D.~Zhu, Phys. Rev.
  Lett. \textbf{127}, 058001 (2021).
\newblock \doi{10.1103/PhysRevLett.127.058001}

\end{thebibliography}
\end{document}